\documentclass[proceedings, preprint]{rmaa}

\usepackage{paralist}

\usepackage{psfrag,color}

\SetYear{}
\SetConfTitle{}

\title{PGMS: to study the Galactic magnetism out of the Galactic plane} 

\author{
  E.~Carretti,\altaffilmark{1} 
  M.~Haverkorn,\altaffilmark{2,3}
  D.~McConnell,\altaffilmark{4}
  G.~Bernardi,\altaffilmark{5} \\
  S.~Cortiglioni,\altaffilmark{6}
  N.M.~McClure-Griffiths,\altaffilmark{4}
  S.~Poppi\altaffilmark{7}
  }

\altaffiltext{1}{INAF -- Istituto di Radioastronomia,
  Via Gobetti 101, 40129 Bologna, Italy
  (e.carretti@ira.inaf.it).}
\altaffiltext{2}{Jansky Fellow, National Radio Astronomy Observatory.}
\altaffiltext{3}{Astronomy Department, University of California, Berkeley, 601 Campbell
                        Hall, Berkeley, CA 94720 (marijke@astro.berkeley.edu).}
\altaffiltext{4}{Australia Telescope National Facility, CSIRO, P.O. Box 76, Epping, NSW
                        1710, Australia (david.mcconnell@csiro.au;  naomi.mcclure-griffiths@csiro.au).}
\altaffiltext{5}{Kapteyn Astronomical Institute, University of Groningen, P.O. Box 800, 9700 AV Groningen, the Netherlands
                         (bernardi@astro.rug.nl).}
\altaffiltext{6}{INAF -- Istituto di Astrofisica Spaziale e Fisica Cosmica Bologna, Via Gobetti 101, 40129 Bologna, Italy 
                         (cortiglioni@iasfbo.inaf.it).}
\altaffiltext{7}{INAF -- Osservatorio Astronomico di Cagliari, Loc. Poggio dei Pini, Strada 54, 09012 Capoterra, Italy
                        (spoppi@ca.astro.it).}

\shortauthor{Carretti et al.}
\shorttitle{PGMS}

\listofauthors{E. Carretti,   M. Haverkorn, D. McConnell,  G. Bernardi, S. Cortiglioni,
                         N.M. McClure-Griffiths, \& S. Poppi}
\indexauthor{Carretti, E.} 
\indexauthor{Haverkorn, M.}
\indexauthor{McConnell, D.}
\indexauthor{Bernardi, G.}
\indexauthor{Cortiglioni, S.}
\indexauthor{McClure-Griffiths, N. M. }
\indexauthor{Poppi, S.}

\abstract{The Parkes Galactic Meridian Survey (PGMS) is a $5^\circ \times 90^\circ$ strip to map 
the polarized synchrotron emission along a Galactic meridian 
from the Galactic plane down to the south Galactic pole. The survey 
is carried out at the Parkes radio telescope
at a frequency of 2.3~GHz with 30 adjacent 8-MHz 
bands which enable Faraday Rotation studies.
The scientific goal is twofold: 
(1)  To probe the Galactic magnetism off the Galactic plane of which little is known so far. 
PGMS gives an insight into the Galactic magnetic field in the thick disc, 
halo, and disc-halo transition;
(2) To study the synchrotron emission as foreground noise of the 
  CMB Polarization, especially for the weak B-Mode which carries the signature 
  of the primordial gravitational wave background left by the 
  Inflation.
PGMS observations have been recently concluded. In this contribution we
present the survey along with first results.}

\resumen{}

\addkeyword{Cosmology: CMB}
\addkeyword{Galaxy: disk}
\addkeyword{Galaxy: halo}
\addkeyword{ISM: magnetic fields}
\addkeyword{polarization}

\begin{document}
% Typeset article header
\maketitle

\section{Introduction}
\label{sec:intro}

The magnetic field is an important component of the Interstellar Medium (ISM) and, 
as in almost any astrophysical context from stars to cosmology, the physics of the Galaxy 
cannot be studied without accounting for it.
 
Observations of other spiral galaxies show that the spiral arms are usually dominated by
a turbulent or tangled component. A weaker coherent field aligned with 
the arms can be  present as well. In the inter-arm regions the regular component is more relevant
and depicts  {\it magnetic arms} with coherent scales up to the size of the disc (e.g. M51, Patrikeev et~al. 2006).
The magnetic field in halos is less clear, instead. This is not only because of the weaker signal, which makes hard to
detect the synchrotron emission, but also of the many different patterns found so far:
from galaxies without evident halo field, to X-shaped fields centred at the galaxy centre (e.g. NGC 891, Krause 2007), 
and up to large almost spherical magnetic halos (e.g. NGC 4631, Hummel \& Dettmar 1990).

Studying other galaxies can help understand the case of the Milky Way, but 
our position internal to the Galaxy makes it harder to figure out how the 
Galactic magnetic field is structured.
 
Probes to study magnetic fields and what is known to date 
about the Galactic field will be briefly reviewed in Sect.~\ref{sec:milkyway}, while
available data of synchrotron polarized emission and requirements for new surveys 
will be discussed in Sect.~\ref{sec:need}. Finally,
the Parkes Galactic Meridian Survey will be presented in Sect.~\ref{sec:pgms}.

\section{The Galactic magnetic field}
\label{sec:milkyway}

\subsection{Probes}
\label{sec:probes}

Three main probes to investigate magnetic fields at radio frequencies are
total synchrotron emission, polarized synchrotron emission, and Rotation Measure (RM)
of pulsars and extragalactic sources.

\begin{enumerate}
  \item
  The synchrotron emission is sensitive to the total magnetic field 
  (ordered + irregular). It dominates the ISM radio emission at frequencies lower than 
  $\sim$5~GHz\footnote{in this frequency range the free-free
  emission can compete only in HII regions in the plane.} and competes with free-free and anomalous dust emission
  up to a few dozen GHz (e.g. de Oliveira-Costa et al. 2008). 
  Several all-sky class surveys are available, covering the range 
  from 20~MHz to 40~GHz (see de Oliveira-Costa et al. 2008 for a mostly complete  
  data set).

  \item
  The polarized synchrotron emission traces the ordered component by three ways. The synchrotron emission
  is intrinsically linearly polarized up to 75\%, but changes in direction of the polarization angle within
  the telescope beam or along the line-of-sight depolarize the signal. Only magnetic fields 
  ordered up to large scales can give a net highly polarized emission. The polarization fraction is thus 
  a measure of the ratio between ordered and  total magnetic field. 

  The polarization angle $\phi$ is rotated by Faraday Rotation (FR) with the wavelength square 
  \begin{equation}
     \Delta\phi = {\rm RM}\,\lambda^2 
  \end{equation}
  where the Rotation Measure RM is
  \begin{equation}
     {\rm RM} = 0.81 \int n_e B_{\parallel}dl
  \end{equation}
  and measures the magnetic field parallel to the line of sight $B_{\parallel}$~[$\mu$G] weighted for the 
  electron density  $n_e$~[cm$^{-3}$] integrated along the path $dl$~[pc] from the source to the observer. 
  Multifrequency observations are required to measure RM and give  the strength of the ordered 
  field along the line-of-sight and its direction (either out- or inward).

  Finally, the synchrotron polarization angle gives the direction of the magnetic field on the
  plane of sky ($B_{\perp}$), once it has been corrected for FR.

  Diffuse polarization data sets are far less complete than those in total emission. Only recently 
  all-sky maps have been completed at 1.4 GHz (Wolleben et al. 2006, Testori et al. 2008) and 
  22.8 GHz (WMAP K-band map: Page et al. 2007),
  but both of them are single frequency and cannot give RM measures. 

  \item 
  RM of Extragalactic Sources (EGS) and pulsars. The former probe $B_{\parallel}$ along the entire line-of-sight through the Galaxy 
 (e.g. Brown et al. 2007). Pulsars, instead, probe out to their position, and, if the position is known
  with sufficient precision, can map the 3D distribution (e.g. Han et al. 2006).
  RM data sets are recently getting richer 
   thanks to surveys like CGPS (Taylor et al. 2003), SGPS (Haverkorn et al. 2006), 
   and Parkes RM pulsar surveys  (e.g. Han et al. 2006, Noutsos et\ al.~2008).
\end{enumerate}

\subsection{Magnetic Field in the Disc}
\begin{figure}[!t]
  \center
  \includegraphics[angle=0, width=1.0\columnwidth]{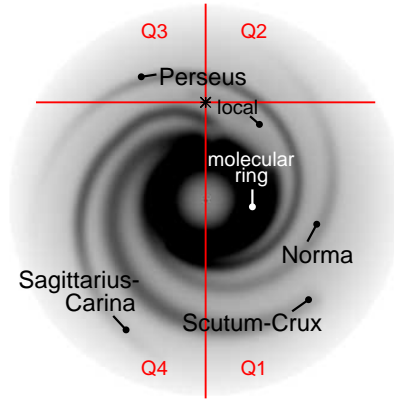}
  \caption{Spiral arms structure of the Milky Way as traced by the electron density model NE2001 (Cordes \& Lazio 2002).
                  Figure courtesy by Jo-Anne Brown.}
  \label{fig:milky}
\end{figure}

Even though the turbulent component dominates the magnetic field in the Galactic plane
making it hard to probe the regular component coherent on large scales, 
the increase of RM data has enabled first studies of  the 3D structure in the disc. 
What seems most favoured is a spiral structure in the disc down to the molecular ring, 
where the field gets a ring shape (e.g. Brown et\ al.~2007, Sun et\ al.~2008). 
The field in the local arm is Clock-Wise (CW) as seen from the north Galactic pole,
while at least one reversal (inversion of the field heading while keeping the same alignment) 
seems to take place toward the inner Galaxy.  
(For a view of the spiral arm structure see Fig.~\ref{fig:milky}, where the electron 
density model by Cordes \& Lazio~2002 is reported.)

%%%%%%%%%%%%%%
%%%%%%%%%%%%%%

The number and location of the reversals is still under debate. 
Han et al. (2006) use pulsar RMs and find evidences 
for reversals at any passage from spiral arms to inter-arm regions, 
with all the arms CW and all inter-arms Counter CW (CCW).  
This would make the Galaxy very peculiar, as all the external galaxies studied so far 
show either no or only one reversal (Beck 2007b). 

Brown et al. (2007) use EGS and find evidence for only one reversal at the inner edge of the Carina--Sagittarius 
arm in the Fourth Quadrant (Q4), with that arm still CW and the Crux--Scutum one CCW. 
That reversal is then kept down to the molecular ring. 
They also find that a second reversal is possible in the Norma arm (CW) but with marginal evidence.
This view is supported by pulsar RM data recently revised by Noutsos et al. (2008) who
find that the field direction in the Carina arm is mainly CW in Q4. This interpretation is supported 
also by Sun et al. (2008), who set up a model of the Galactic magnetic field accounting for 
several types of data sets (RMs, total and polarized emission
surveys, and others). They find that an Axi-Symmetric Spiral (ASS) structure
with one reversal at the inner edge of the Carina--Sagittarius arm best fit the data. 
They also found that the data are equally well fitted by an ASS model plus a ring structure 
in the inner Galaxy, with CCW field in the ring and CW in the ASS section.

It is worth noticing that the CW direction of the Carina--Sagittarius arm in Q4 
is in contrast with its CCW direction in Q1, which has strong evidences 
(e.g. Thomson \& Nelson 1980). 
Therefore, the Carina--Sagittarius arm should present a field reversal internal to the arm 
somewhere close to the Q1-Q4 border.
In fact, the arm is {\it distorted} with respect a {\it normal} spiral arm, also with
a misalignment between spiral arm and magnetic field direction (Beck 2007a). 
This would be compatible with a RING model for the inner Galaxy with a field reversal at the
outer edge of the ring. The field reversal internal  the Carina-Sagittarius arm would occur 
where the arm passes through the outer edge of the ring. 
Besides to be one of the best fits  of Sun et al. (2008), the inner RING model has 
been first proposed by Vall\'ee (2005).

In any case, it is clear that, despite the recent big observational efforts, the framework is not yet fully 
understood and more data are necessary to disentangle among the different interpretations.

\subsection{Magnetic Field in the Galactic Halo}
The data available for the Galactic halo field are far fewer than those in the disc 
and models are even  less constrained. 
Our knowledge is basically based on RM data at mid and 
high Galactic latitudes, which, however, are coarsely and irregularly sampled. 
Han (2002)  finds that RMs values are asymmetric both with respect to the plane and the Galactic centre,
which is compatible with a field generated by an $\alpha$--$\Omega$ dynamo model of A0-mode: 
 two toroidal fields above and under the Galactic plane  antisymmetric across the plane (sign reversal).
 However, diffuse polarization maps at 1.4 and 22.8~GHz 
 clearly show very large  structures in that region extending from the plane up to high latitudes, which likely
 are very local emissions. Contamination of RM data by local {\it anomalies} is therefore possible, 
 jeopardising the interpretation as large scale field. 
 Sun et al. (2008) found  that their model cannot yet constrain the halo field essentially for lack of data,
 and its parameters remain open.
It is also worth mentioning that there is no observation of external galaxy 
with large coherent dynamo pattern yet, and the most extended fields are likely
sustained by wind outflows generated by the underlying active star-forming disc 
(Beck 2007b).

In conclusion, very little is known about the Galactic halo field so far 
and present data cannot significantly
constrain models. Even more than for the disc, far more data are necessary to understand
the magnetic field structure of the halo.

\section{Polarized Diffuse Emission: status and survey needs}
\label{sec:need}

%%%%%%%%%%%%%%
\begin{figure}[!t]
  \center
  \includegraphics[angle=90, width=1.0\columnwidth]{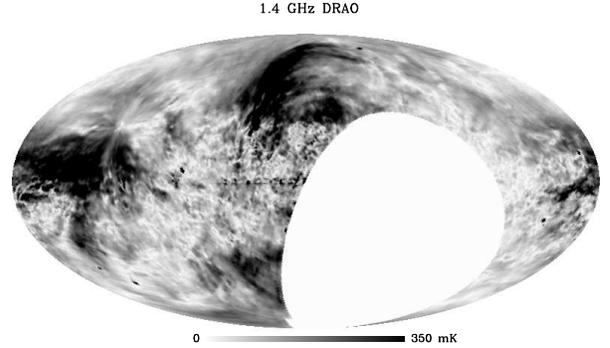}
  \caption{Polarized intensity emission at 1.4~GHz  of the northern sky (DRAO survey, Wolleben et al. 2006).
                  The map is a Mollweide projection in Galactic coordinates using the HEALPix pixelation 
                  (G{\'o}rski et al. 2005). It is centred at Galactic Centre and longitude increases leftward.}
  \label{fig:drao}
\end{figure}
%%%%%%%%%%%%%%
%%%%%%%%%%%%%%
\begin{figure}[!t]
  \center
  \includegraphics[angle=90, width=1.0\columnwidth]{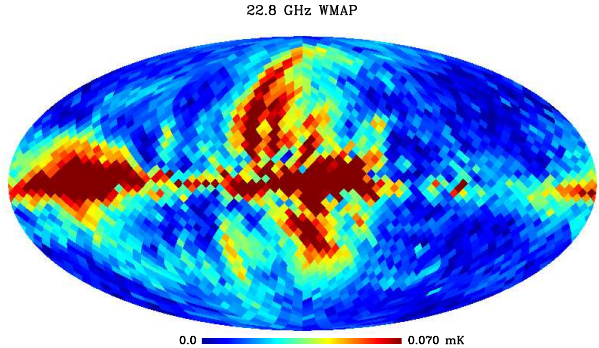}
  \caption{Polarized intensity emission at 22.8~GHz (WMAP survey, Page et~al. 2007).
                  Data are binned in $\sim 4^\circ$ pixels.
                  Projection and pixelation are as for Fig.~\ref{fig:drao}}
  \label{fig:wmap}
\end{figure}
%%%%%%%%%%%%%%

Diffuse polarized emission surveys carry  much information about magnetic fields 
(Sect.~\ref{sec:probes}). The most important available data are the two all-sky surveys 
which have been recently completed at 1.4 and 22.8 GHz. 

The 1.4 GHz survey  is the combination of a Northern (Fig.~\ref{fig:drao}) and
Southern section (Wolleben et al. 2006, Testori et al. 2008, respectively).
It has been the first all-sky coverage in polarization allowing the first look at the 
{\it ordered} magnetized component of the Galaxy. 
The emission is affected by FR with strong depolarization in the disc up 
to latitude $b\sim |30^\circ|$ (Wolleben et al. 2006) and alterations up to  
$b\sim |50^\circ|$ (Carretti et al. 2005).
However, the survey is single band and is not self-sufficient to measure RMs and give
intrinsic polarization angles. 

\begin{figure*}[!t]
  \center
  \includegraphics[angle=90, width=0.8\textwidth]{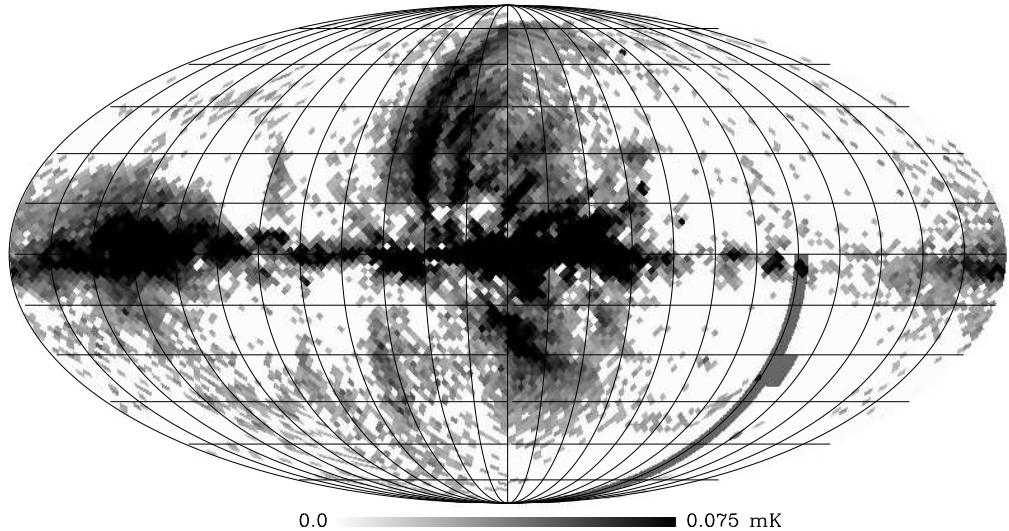}
  \caption{The PGMS strip (grey) plotted on the WMAP polarized emission map at 22.8 GHz (Page et al. 2007). 
                  The map is in Galactic coordinates with $l=0^\circ$ at centre and longitude increasing leftward. 
                  The PGMS strip is centred at $l=254^\circ$. The enlargement at $b=-35^\circ$ is also shown.
                  Data are binned in $\sim2^\circ$ pixels and pixels with S/N $<$ 3 have been blanked (white).
                  }
  \label{fig:pgms}
\end{figure*}

The 22.8~GHz all-sky map by WMAP (Page et al. 2007) is the first map at microwave frequencies 
where FR effects are negligible (Fig.~\ref{fig:wmap}). It shows large local structures up to high
latitudes. However, it has not sufficient sensitivity to detect the signal out of these local structures 
both in the disc and at high latitudes (Fig.~\ref{fig:pgms}; see also Carretti et al. 2006), 
which prevents the investigation in regions clear by local deviations.

%%%%%%%%%%%%%%
%%%%%%%%%%%%%%

In spite of these recent efforts, the polarized diffuse emission 
is not yet sufficiently observed and more data are necessary.
The features that new surveys should have can read as follows:
\begin{description}

  \item{--} observations at radio frequencies, where the synchrotron emission dominates;
  
  \item{--} at frequency $\nu > 1.4$~GHz, to prevent depolarization;

  \item{--} at frequency not too high, to benefit from a good $S/N$;

  \item{--} many frequency channels, to enable RM computation;
  
  \item{--} observations at all latitudes from the Galactic plane to the pole, to access all environments;
  
  \item{--} observations of regions clear from strong local structures, to prevent contamination from
                local anomalies.

\end{description}

\section{Parkes Galactic Meridian Survey}
\label{sec:pgms}

The Parkes Galactic Meridian Survey (PGMS) is a project 
to survey the diffuse polarized emission along a Galactic meridian at a frequency of 2.3~GHz. 
Carried out at the Parkes radio telescope, 
it consists of a $5^\circ$ wide and  $90^\circ$ long 
strip to cover all the southern meridian at longitude $l=254^\circ$ 
from the Galactic plane down to the south Galactic pole (Fig.~\ref{fig:pgms}). 
It also includes a $10^\circ \times10^\circ$ enlargement centred at $b=-35^\circ$.
Main features are reported in Tab.~\ref{tab:pgms}.

\begin{table*}[!t]\centering
  \setlength{\tabnotewidth}{0.9\textwidth}
 \tablecols{8}
  \setlength{\tabcolsep}{1.0\tabcolsep}
  \caption{PGMS main parameters\tabnotemark{a}} \label{tab:pgms}
  \begin{tabular}{cccccccc}
    \toprule
   $\nu$    & BW    & $N_{\rm ch}$ ~~&  FWHM &  $l_0$     &  $b$-range   & Area size &         $\sigma_{Q,U}$ \\
    \midrule
  2300~MHz~~ & ~~240~MHz~~ & ~~30~~ &~~8.9'~~& ~~$254^{\circ}$~ & $[0^\circ, -90^\circ]$ &
  ~~~~$5.0^{\circ}\times 90.0^{\circ}$~~ &   ~~0.3 mK \\
    \bottomrule
    \tabnotetext{a}{$\nu$ is the central frequency, BW 
the bandwidth, $N_{\rm ch}$ the number of frequency channels, FWHM the beam size, $l_0$ 
the central strip meridian, and $\sigma_{Q,U}$ the sensitivity of Stokes $Q$ and $U$ in a 
beam-size pixel. The area size is also reported. }
  \end{tabular}
\end{table*}

Multifrequency observations have been conducted in 30~adjacent channels 8~MHz wide to enable RM measurements.
The expected  performances are: 
\begin{description}
   \item{--} In the disc, emission and RM are sufficiently high that RM can be measured with 
                 PGMS data alone using its 30 frequency channels. The expected sensitivity is 
                 $\sim 10$~rad/m$^2$ on a scale of 9~arcmin, sufficient for disc RM values which are
                 hundreds rad/m$^2$.
                 
  \item{--} In the halo, RM can be measured in combination with the 1.4~GHz data
                with an expected sensitivity of $\sim2$~rad/m$^2$  at the 36~arcmin resolution of the 
                1.4~GHz survey. 
\end{description}

The central frequency of 2.3~GHz has been chosen as a trade-off to be high enough to prevent Faraday 
depolarization (apart from few degrees around the Galactic plane, possibly), 
and sufficiently low to ensure the appropriate S/N.

The selected meridian goes through lowest emission regions clear from large local structures,
as visible in Fig.~\ref{fig:pgms} where the strip is projected onto the WMAP map. 
This enables PGMS to probe the Galactic field without significant local deviations.

PGMS fulfils the requirements listed in Sect.~\ref{sec:need} and the RM performances
allows a significant improvement of mapping the Galactic field both in resolution and sensitivity. 
Although PGMS is not all-sky, its exploration of all the latitudes without local anomalies is expected 
to give new insights into the field in the halo, disc and at the disc-halo transition.

The structure and strength of the field in the halo are important
to distinguish among dynamo models (especially the vertical component) and the high RM sensitivity of PGMS
is expected to give important new information. 

The 1.4~GHz data are strongly depolarized in the disc at latitudes $|b|< 30^\circ$, 
which, thus, is mostly unexplored in diffuse polarized emission.
Both the higher frequency and resolution are expected to enable PGMS to detect the signal also in the disc.

Finally, the 1.4~GHz data show a sudden transition at  $|b|\sim30^\circ$. PGMS aims at  
exploring it to monitor the transition from coherent to turbulent field.

The other main goal of PGMS is the study of the synchrotron emission as foreground noise for CMB 
Polarization observations (CMBP). The CMBP is predicted to carry the signature 
of the primordial gravitational wave background emitted by the Inflation, 
which opens the possibility to directly probe that event of the primeval Universe. 
However, the signal is weak and contaminated
by the synchrotron emission even at high Galactic latitude (Page et~al.~2007). To clean 
it is  mandatory, which requires high sensitivity mapping. 
On the other hand, some areas of the sky seems cleaner (cfr. Carretti et~al.~2006 for a review). 
The areas observed so far are too few and small to extend their results to larger sky areas, however.  
PGMS goes through lowest emission regions and can give statistically robust  estimates 
of the contamination level of these best areas.  
This is especially important for ground-based and balloon-borne experiments, like CLOVER (Taylor~2006)
and EBEX (Oxley et~al.~2004),  which usually observe in small sky areas and 
can benefit of particularly clean regions.

%%%%%%%%%%%%%%
\begin{figure}[!t]
  \center
  \includegraphics[angle=0, width=1.0\columnwidth]{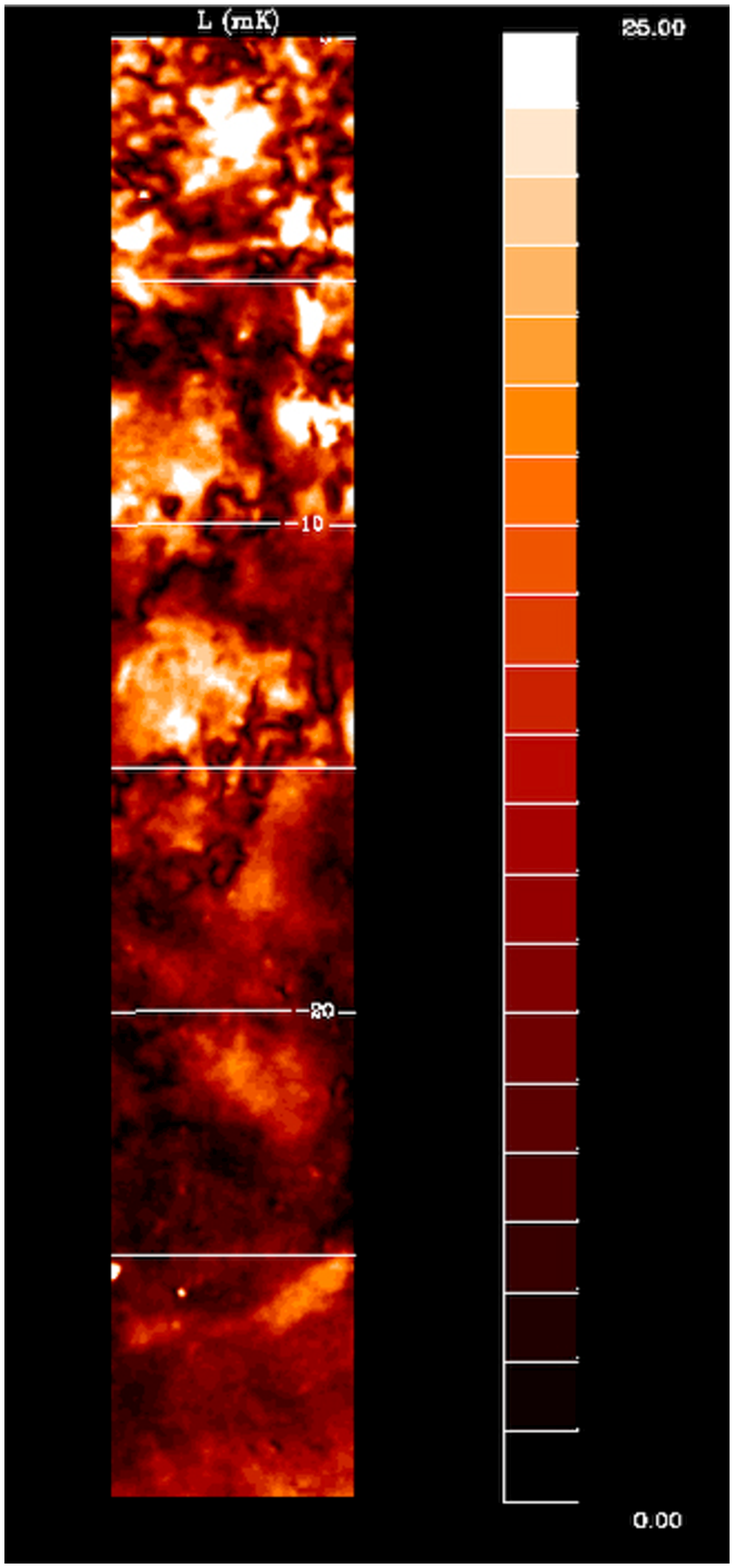}
  \caption{Polarized intensity  $L = \sqrt{Q^2+U^2}$ [mK] of the PGMS fields in the Galactic disc 
                  at latitude $b=[0^\circ, -30^\circ]$. Isolatitude
                  contours are spaced by $5^\circ$.}
  \label{fig:discmap}
\end{figure}
%%%%%%%%%%%%%%
%%%%%%%%%%%%%%
\begin{figure}[!t]
  \center
  \includegraphics[angle=0, width=1.0\columnwidth]{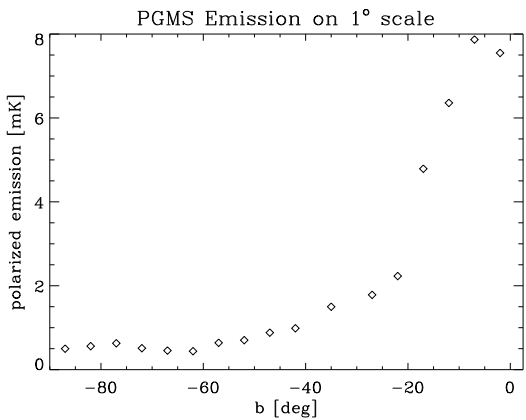}
  \caption{Polarized emission on $1^\circ$ scale of the PGMS fields plotted versus Galactic latitude $b$. }
  \label{fig:linvslat}
\end{figure}
%%%%%%%%%%%%%%

PGMS observations have been completed in September 2007 after 4 runs spread in 2 years.
The strip to observe has been divided in $5^\circ \times 5^\circ$ fields, except the enlargement centred 
on $b=-35^\circ$ which is  $10^\circ \times 10^\circ$. 
Each field includes a $1^\circ$ enlargement along $b$ at the north edge, 
so that the actual area of each field is $5^\circ \times 6^\circ$ with a 
$5^\circ \times 1^\circ$ overlap with the next northern field.
The final sensitivity per beam-size pixel is $\sim0.3$~mK 
for high latitude areas ($|b| > 30^\circ$) and $\sim0.5$~mK for those of the disc ($|b| < 30^\circ$).

Figure~\ref{fig:discmap} shows the polarized intensity of the disc fields. 
The strongest emission is near the Galactic plane and keeps 
high up to $|b| = 15^\circ$--$20^\circ$, where it starts to drop down. This behaviour can be better appreciated in 
Figure~\ref{fig:linvslat}, where the emission on $1^\circ$ scale 
versus latitude is plotted.  A transition at $|b| = 15^\circ$--$20^\circ$ is evident, while
the decline continues at higher latitudes up to $|b| \sim 40^\circ$ where the emission
sets on low levels up to the south Galactic pole. This behaviour differs from the 1.4~GHz data, 
where the disc emission is depolarized and weaker than at high latitude. 
 
 In addition, the 2.3~GHz emission looks mottled from the plane up to $|b| = 6^\circ$--$7^\circ$, but
 then becomes smooth. This supports that FR effects are marginal even in the disc at this frequency
  and affect the emission only within a few degrees around the Galactic plane. Figure~\ref{fig:linvslat} also shows a slight depletion in the first field at
 $b=[0^\circ, -5^\circ]$.
 
 This behaviour supports that 2.3~GHz (and 9' resolution) is a frequency sufficiently high to show the polarized 
 emission in most of the disc with no (or marginal) Faraday depolarization. 
 Faraday rotation modulation is evident only in the very few degrees close to 
 the plane, but it is not yet clear whether there is depolarization and, if yes, how much it is. 
 The on-going multifrequency analysis should help us work it out especially through the RM analysis. 
 
 About the CMB foreground analysis, the low Faraday effects ensure safe extrapolations 
 to the CMB frequencies at high latitudes. 
 Computation of the angular power spectra is on-going and a full analysis will be soon published. 
 Here we want only to notice the presence of a best clean area
 at $b=[-75^\circ, -60^\circ]$ which looks of high interest for sub-orbital experiments.
 
The PGMS data analysis is in progress and results will soon be published, 
but its {\it successor} has started already. 
The S-PASS (S-band Polarization All Sky Survey, Carretti et al. 2007) 
is aimed at  surveying the whole southern sky with the Parkes
telescope and with the same set up of PGMS, with obvious advantages of full
spatial characterization of the polarized diffuse emission in this band.
S-PASS is 30\% complete to date (April 2008) and is expected to be finished by 2009.

\begin{acknowledgements}
This work is partly supported by the ASI project COFIS.
MH  acknowledges support from the National Radio Astronomy Observatory (NRAO), 
which is operated by Associated Universities Inc., under cooperative agreement with the 
National Science Foundation.
Part of this work is based on observations taken with
the Parkes Radio telescope, which is part of the Australia Telescope,
funded by the Commonwealth of Australia for operation
as a National Facility managed by CSIRO.
Some of the results in this paper have been derived using the 
HEALPix package (http://healpix.jpl.nasa.gov, G{\'o}rski et~al.~2005).
The Dominion Radio Astrophysical Observatory is operated as a national Facility by the National Research Council Canada.
We acknowledge the use of WMAP data, and the Legacy Archive for Microwave Background
Data Analysis (LAMBDA). Support for LAMBDA is provided by the NASA Office of Space Science.
\end{acknowledgements}

\end{document}